\documentstyle[aps,graphicx,epsf,amssymb,eqsecnum]{revtex}
\begin{document}

\draft

\title{Wave Number of Maximal Growth in Viscous Magnetic Fluids of Arbitrary
Depth}
\author{Adrian Lange}
\address{Universit\"at Magdeburg, Institut f\"ur Theoretische
Physik, Universit\"atsplatz 2, D-39106 Magdeburg, Germany}
\author{Bert Reimann and Reinhard Richter}
\address{Universit\"at Bayreuth, Physikalisches Institut, Experimentalphysik V,
D-95440 Bayreuth, Germany}
\date{\today}
\maketitle
\begin{abstract}
An analytical method within the frame of linear stability theory
is presented for the normal field instability
in magnetic fluids. It allows to calculate the maximal growth rate
and the corresponding wave number for any combination of thickness
and viscosity of the fluid.
Applying this method to magnetic fluids of finite depth, these results are
quantitatively compared to the wave number of the transient pattern
observed experimentally after a jump--like increase of the field.
The wave number grows linearly with increasing induction where the
theoretical and the experimental data agree well.
Thereby a long-standing controversy about the behaviour of
the wave number above the critical magnetic field is tackled.
\end{abstract}

\section{Introduction}

Spontaneous pattern formation from a homogeneous ground state has been studied
extensively in many nonlinear dissipative systems. Among these systems magnetic
fluids have experienced a renewed interest in recent years due to their
technological importance \cite{handbook}. The most striking phenomenon of
pattern formation in magnetic fluids is the Rosensweig or normal field
instability \cite{cowley,rosensweig,gailitis,bashtovoi}.
Above a threshold $B_c$ of the induction
the initially flat surface exhibits a stationary hexagonal pattern of peaks.
Typically, patterns are characterized by a wave vector ${\bf q}$ whose absolute
value gives the wave number $q=|{\bf q}|$. In contrast to many other systems, a
comprehensive {\it quantitative} theoretical and experimental analysis of the
dependence of the wave number on the strength of the magnetic field is lacking
for the normal field instability. There are few but contradictory experimental
observations. In experiments where the field is increased continuously, there
are reports about constant \cite{cowley,bacri84} as well as about varying wave
numbers \cite{mahrdiss} as the induction is increased beyond the critical value
$B_c$. Notably, all these observations are of entirely {\it qualitative}
character \cite{wesfreid}.

A first theoretical analysis leading to constant wave numbers of maximal growth
was presented in \cite{salin93}. The general dispersion relation for surface waves
on a magnetic fluid of infinite thickness was analysed for two asymptotic regimes:
for the inviscid regime and for the viscous-dominated regime. The main result
for the latter regime was that taking into account viscous effects the wave number of
maximal growth is the same at and {\it beyond} the critical induction.
As will be shown below, this argument is rather misleading because realistic fluid
properties are not covered by such an asymptotic analysis. The two asymptotic regimes
in \cite{salin93} were combined with very thin as well as very thick layers of
magnetic fluid and the resulting four regimes were analysed in \cite{abou97}.
In three regimes a non-constant wave number of maximal growth was found.

All qualitative observations in \cite{cowley,bacri84,mahrdiss} are referring to the
final arrangement of peaks. The final stable pattern, resulting from
{\em nonlinear} interactions, does not generally correspond to the most unstable
{\em linear} pattern. Such a pattern should grow with the maximal growth rate and
should display the corresponding wave number. Since both quantities are calculated
by the linear theory, the most unstable linear pattern has to be detected and
measured experimentally for a meaningful comparison between theory and experiment.
No measurements of the most linear unstable pattern have yet been undertaken.

Motivated by this puzzling situation, the paper presents a {\it quantitative}
theoretical analysis of the wave number with maximal growth rate
for any combination of fluid parameters. Experimental
measurements of the most linear unstable pattern are conducted and the
data compared with the theoretical results. The system and the
relevant equations of the problem are displayed in the next Section.
Based on the dispersion relation from a linear stability analysis, an
analytical method is presented to calculate $q_m$ and the
maximal growth rate $\omega_m$ for any combination of material
parameters. The details of the method are explained for a magnetic fluid of
infinite thickness and the results are compared with previous asymptotic
results
(Sec.\ III). The method is also applied to magnetic fluids of finite thickness
(Sec.\ IV), which allows a quantitative comparison with the experimental data
(Sec.\ V).
In the final section the results are summarized and further prospects
are outlined.

\section{System and Equations of the Problem}

A horizontally unbounded layer of an incompressible, nonconducting,
and viscous magnetic fluid of thickness $h$ and constant density
$\rho$ is considered. The fluid is bounded from below ($z=-h$) by
the bottom of a container made of a magnetically
impermeable material and has a free surface described by $z=\zeta (x, y, t)$
with air above. The electrically insulating
fluid justifies the stationary form of the Maxwell equations which reduce
to the Laplace equation for the magnetic potentials $\Phi^{(i)}$ in each of
the three different regions. (Upper indices denote the considered medium:
$1$ air, $2$ magnetic fluid, and $3$ container.) It is assumed that the
magnetization ${\bf M}^{(2)}$ of the magnetic fluid depends linearly on the
applied magnetic field ${\bf H}^{(2)}$, ${\bf M}^{(2)} =(\mu_r -1){\bf
H}^{(2)}$, where
$\mu_r$ is the relative permeability of the fluid. The system is
governed by the equation of continuity and the Navier-Stokes equations
for the magnetic fluid
\begin{eqnarray}
  \label{eq:1}
  {\rm div}\, {\bf v} & = & 0 \;,\\
  \label{eq:2}
  \partial_t {\bf v} + \left( {\bf v}\, {\rm grad}\right){\bf v} & = & -{1\over
  \rho}\,
  {\rm grad}\left( p +p_s \right) +\nu\Delta {\bf v} + {\bf g}\; ,
\end{eqnarray}
and the Laplace equation in each medium
\begin{equation}
  \label{eq:3}
  \Delta\Phi ^{(i)} = 0\; .
\end{equation}
The quantities without an upper index are referring to the magnetic fluid with the
velocity field ${\bf v} =(u, v, w)$, the kinematic viscosity
$\nu$, the pressure $p$ and the acceleration due to gravity ${\bf g}$. The
first three terms on
the right-hand side of Eq.~(\ref{eq:2}) result from $(1/\rho){\rm
div}\,\tensor{T}^{(2)}$
where the components of the stress tensor $\tensor{T}^{(2)}$ read
\cite{rosensweig}
\begin{equation}
  \label{eq:4}
  T_{ij}^{(2)} = \left\{ -p -\mu_0\int_0^{H}\!\!\!\left( M -
  \rho\,\partial_\rho M\right) dH' - \mu_0{H^2\over 2}\right\}\delta_{ij} +H_i
  B_j
  +\rho\,\nu\left(\partial_i v_j +\partial_j v_i \right)\, .
\end{equation}
The magnetostrictive pressure is given by $p_s = - \mu_0\int_0^{H}\!\!\!\rho\,
\partial_\rho M dH'$.
$M$, $H$, and $B$ denote the absolute value of the magnetization, the magnetic
field and the induction
${\bf B}$ in the fluid. The governing
equations have to be supplemented by the appropriate boundary conditions which
are
the continuity of the normal (tangential) component of the induction (magnetic
field) at the top
and bottom interface
\begin{eqnarray}
  \label{eq:5}
  &&{\bf n}\cdot\left( {\bf B}^{(1)}-{\bf B}^{(2)}\right) =0\qquad
  {\bf n}\times\left( {\bf H}^{(1)}-{\bf H}^{(2)}\right) =0
  \qquad\qquad {\rm at~} z=\zeta\\
  \label{eq:6}
  &&{\bf n}\cdot\left( {\bf B}^{(2)}-{\bf B}^{(3)}\right) =0\qquad
  {\bf n}\times\left( {\bf H}^{(2)}-{\bf H}^{(3)}\right) =0
  \qquad\qquad {\rm at~} z= -h\, ,
\end{eqnarray}
the no-slip condition for the velocity at the bottom of the container
\begin{equation}
  \label{eq:7}
  {\bf v}=\partial_z w=0\qquad\qquad  {\rm at~} z= -h\, ,
\end{equation}
the kinematic boundary condition at the free surface
\begin{equation}
  \label{eq:8}
  w =\partial_t \zeta +({\bf v}\,{\rm grad})\,\zeta\qquad\qquad  {\rm at~} z=
  \zeta\, ,
\end{equation}
and the continuity of the stress tensor across the free surface
\begin{equation}
  \label{eq:9}
  n_i\left( T_{ij}^{(1)} - T_{ij}^{(2)}\right) = -\sigma K n_j\qquad\qquad  {\rm
  at~} z= \zeta\, .
\end{equation}
The surface tension between the magnetic fluid and air is denoted by $\sigma$,
the
curvature of the surface by $K={\rm div}\, {\bf n}$, and the unit vector normal
to the surface by
\begin{equation}
  \label{eq:10}
  {\bf n} = {{\rm grad}\left[ z-\zeta (x,y,t)\right]\over |{\rm grad}\left[ z-
  \zeta (x,y,t)\right]|} ={(-\partial_x\zeta , -\partial_y\zeta , 1)\over
  \sqrt{1+(\partial_x\zeta)^2 + (\partial_y\zeta)^2}}\, .
\end{equation}
Since the density of air can be neglected with respect to the density of the
magnetic
fluid and $M^{(1)}=0$ holds, Eq.~(\ref{eq:9}) reduces to
\begin{equation}
  \label{eq:11}
  n_j\left\{ -p^{(1)}+ p + \mu_0\int_0^H\!\!\! M dH' + p_s  +{\mu_0\over
  2}M_n^2\right\}
  -\rho\,\nu\, n_i\left\{ \partial_i v_j + \partial_j v_i\right\} = -\sigma K
  n_j\, ,
\end{equation}
where $p^{(1)}$ is the atmospheric pressure above the fluid layer. In a linear
stability analysis all small disturbances from the basic state
are analysed into normal modes, i.e., they are proportional to
$\exp [-i(\omega\,t -\vec q\,\vec r)]$. If $\Im (\omega) >0$, initially small
undulations will grow exponentially and the originally horizontal surface is
unstable.
Due to this relation it has been established to denote $\omega$
as growth rate which is in fact true only for its imaginary part.

Following the standard procedure the linear stability analysis leads to the
dispersion relation \cite{abou97,weilepp,mueller98}
(all formulas in the references are equivalent to each other)
\begin{eqnarray}
  \nonumber
  0 & = &{\nu^2\over \tilde q\coth(\tilde q h)- q\coth (qh)}\left\{
    \tilde q\left[ 4q^4 +(q^2+\tilde q^2)^2\right]\coth(\tilde q h)
    -q\left[ 4q^2\tilde q^2+(q^2 +\tilde q^2)^2\right]\tanh (qh)
    -{4q^2\tilde q (q^2+\tilde q^2)\over \cosh (qh) \sinh (\tilde q
    h)}\right\}\\
   && +\tanh (qh) \left[ gq +{\sigma\over \rho}q^3-{\mu_0 \mu_r M^2\over \rho}
   \Lambda (qh)\,q^2\right]\, ,
  \label{eq:12}
\end{eqnarray}
where $\mu_0$ is the permeability of free space, $\tilde q =\sqrt{q^2
-i\omega/\nu}$, and
\begin{equation}
  \label{eq:13}
  \Lambda (qh) = {{\rm e}^{qh} (1+\mu_r) + {\rm e}^{-qh}(1-\mu_r)\over
  {\rm e}^{qh}(1+\mu_r)^2 - {\rm e}^{-qh}(1-\mu_r)^2}\, .
\end{equation}
The condition of marginal stability, $\omega =0$, defines the threshold where
$\omega$ changes its sign
and therefore the normal field or Rosensweig instability appears. With $\omega
=0$ one
obtains from Eq.~(\ref{eq:12})
\begin{equation}
  \label{eq:14}
  {\rho\, g\over \sigma} + q^2 -{(\mu_r -1)^2 B^2\over
  \mu_0\,\mu_r\,\sigma}\,q\,\Lambda (qh) = 0\, .
\end{equation}
In the limit of an infinitely thick ($h\rightarrow\infty$) or an infinitely
thin
($h\rightarrow 0$) layer, respectively, the critical inductions are
\begin{eqnarray}
  \label{eq:15}
  B_{c,\infty}^2 &=& {2\mu_0\,\mu_r (\mu_r +1)\sqrt{\rho\,\sigma\,g}\over
  (\mu_r -1)^2}\\
  \label{eq:16}
  B_{c,0}^2 &=& {4\mu_0\,\mu_r^2 \sqrt{\rho\,\sigma\,g}\over (\mu_r -1)^2}\, ,
\end{eqnarray}
whereas in both limits the critical wave number is equal to
\begin{equation}
  \label{eq:17}
  q_c = q_{c,\infty}=q_{c,0}=\sqrt{{\rho\, g\over \sigma}}\, .
\end{equation}
The critical values apply to both viscous and inviscid magnetic fluids due to
the static character of
the instability. Based on the dispersion relation (\ref{eq:12})
the details of the proposed new method are presented exemplarily for a magnetic
fluid of
infinite thickness in the next section.

\section{Infinite Layer of Magnetic Fluid}
\begin{figure}[htbp]
  \begin{center}
    \includegraphics[scale=0.45]{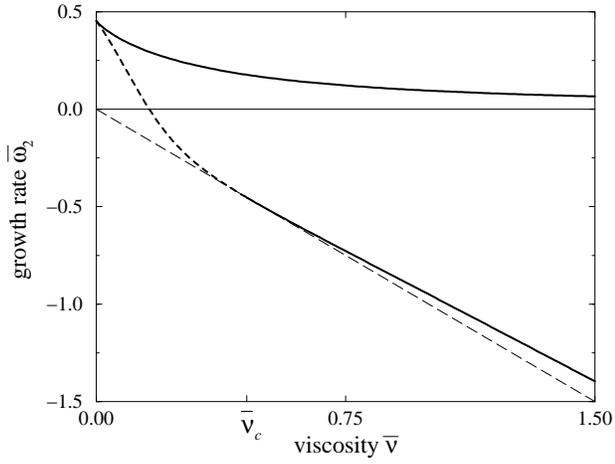}
    \caption{Purely imaginary growth rate $\bar\omega =i\,\bar\omega_2$ as
    function
    of the viscosity for $\bar q=1$. For $k=0$ (solid lines) positive values of
    $\bar\omega_2$ exist
    for all viscosities whereas negative ones exist only above a critical
    viscosity
    $\bar\nu_c =0.453$. For $k=1$ (dashed line) the upper bound for $\bar\omega
    =i\,\bar\omega_2$
    is given by $\bar\nu_c$. The long-dashed line indicates the condition
    $\bar\nu\bar q^2 =|\bar\omega |$ for $\bar\omega_2 <0$.}
    \label{fig:1}
  \end{center}
\end{figure}

The starting point of the analysis is the determination of the parameters for
which the
dispersion relation (\ref{eq:12}) for an infinitely thick layer \cite{salin93}
\begin{equation}
  \label{eq:41}
  \left( 1-{i\omega\over 2\nu q^2}\right)^2 +
  {1\over 4\rho \nu^2 q^4}\left[ \rho g q + \sigma q^3 - {(\mu_r -1)^2\over
  (\mu_r +1)
  \mu_0 \mu_r} B^2 q^2\right] = \sqrt{ 1 -{i\omega \over \nu q^2} }
\end{equation}
has solutions of purely imaginary growth rates. Such growth rates characterize
the viscous-dominated regime described by $q\delta\gg 1$ \cite{salin93},
where $\delta =\sqrt{2\nu /\omega}$ denotes the viscous depth \cite{comment1}.
For $\omega$ the polar representation
$\omega =\omega_1 + i\omega_2=|\omega | \left(\cos\varphi_0
+i \sin\varphi_0\right)$ is chosen with
\begin{equation}
  \label{eq:42}
   \varphi_0=\arctan {\omega_2\over \omega_1}
    +\left\{ \begin{array}{c@{\quad }c}
                     {\displaystyle{0}} & {\rm if~}n\geq 0{\rm
                     ~and~}d>0\phantom{\, ,}\\
                     {\displaystyle{\pi}} & {\rm if~}n\geq 0{\rm
                     ~and~}d<0\phantom{\, ,}\\
                     {\displaystyle{\pi/2}} & {\rm if~}n\geq 0{\rm
                     ~and~}d=0\phantom{\, ,}\\
                     {\displaystyle{2\pi}} & {\rm if~}n <0{\rm
                     ~and~}d>0\phantom{\, ,}\\
                     {\displaystyle{\pi}} & {\rm if~}n <0{\rm
                     ~and~}d<0\phantom{\, ,}\\
                     {\displaystyle{{3\pi/2}}} & {\rm if~}n<0{\rm ~and~}d=0\, ,
                     \end{array} \right.
\end{equation}
where $n$ ($d$) denotes the numerator (denominator) of the argument of
$\arctan$.
Dimensionless quantities were introduced for all lengths, the induction, the
time, and the viscosity
\begin{eqnarray}
  \label{eq:43}
  \bar l &=& q_c\; l\hskip 1.65 cm
  \qquad\qquad\>\,\bar B = {B\over B_{c,\infty}}\\
  \label{eq:44}
  \bar t &=& {g^{3/4}\rho^{1/4}\over \sigma^{1/4}}\; t ={t\over t_c}
  \qquad\qquad \bar \nu ={ g^{1/4} \rho^{3/4} \over \sigma^{3/4} }\; \nu\,
\end{eqnarray}
where $t_c$ is the so-called capillary time.
The real and imaginary part of (\ref{eq:41}) now read
\begin{eqnarray}
  \nonumber
  \bar\nu ^2 -{|\bar\omega |^2\left(\cos^2 \varphi_0 -\sin^2\varphi_0\right)
  \over 4 \bar q\, ^4}
  +{\bar \nu |\bar\omega | \sin\varphi_0\over \bar q\, ^2}
  &+&{\bar q +\bar q\, ^3 -2 \bar B^2 \bar q\, ^2 \over 4 \bar q\, ^4}\\
  \label{eq:46}
  &=&\bar\nu^2\sqrt[\scriptstyle{4}]{\left(1+{|\bar\omega |\sin\varphi_0 \over
  \bar\nu \bar q\, ^2}\right)^2
   +{|\bar\omega |^2 \cos^2 \varphi_0\over \bar\nu^2 \bar q\, ^4}}\;
   \cos\left({\psi +2 k\pi\over 2}\right) \quad k=0,1 \\
  \label{eq:47}
  -{|\bar\omega|^2\sin\varphi_0\cos\varphi_0\over 2\bar q\, ^4}-
   {\bar\nu|\bar\omega |\cos\varphi_0\over \bar q\, ^2}
  &=&\bar\nu^2\sqrt[\scriptstyle{4}]{\left(1+{|\bar\omega |\sin\varphi_0 \over
  \bar\nu \bar q\, ^2}\right)^2
   +{|\bar\omega |^2 \cos^2 \varphi_0\over \bar\nu^2 \bar q\, ^4}}\;
   \sin\left({\psi +2 k\pi\over 2}\right)\quad k=0,1\; ,
\end{eqnarray}
where
\begin{equation}
  \label{eq:48}
  \psi =\arctan\,{-|\bar\omega |\cos\varphi_0\over \bar\nu\bar q^2 +
        |\bar\omega |\sin\varphi_0}
        +{\rm const.}
\end{equation}
and $k$ distinguishes between the two possible values of the complex root.
The value of the constant in (\ref{eq:48}) follows the rules of (\ref{eq:42}).
For a purely imaginary growth rate, $\bar\omega= i\bar\omega_2$, $\varphi_0$
can take
only the two values $\pi/2$ ($\bar\omega_2>0$) and $3\pi/2$
($\bar\omega_2<0$). In the former case
Eq.~(\ref{eq:47}) is always fulfilled, whereas in the latter case
Eq.~(\ref{eq:47}) holds
only if $\bar\nu\bar q^2 > |\bar\omega |$. For $\bar\omega= i\bar\omega_2$
Eq.~(\ref{eq:46}) reduces to
\begin{equation}
  \label{eq:49}
  f_\pm (\bar q , |\bar\omega |;{\bar\nu ,\bar B}):=
  \left(\bar\nu \pm {|\bar\omega |\over 2\bar q\, ^2}\right)^2
  +{\bar q +\bar q\, ^3 -2 \bar B^2 \bar q\, ^2\over 4 \bar q\, ^4}
  -\bar\nu^2\sqrt[\scriptstyle{4}]{\left( 1 \pm {|\bar\omega |\over \bar\nu\bar
  q\, ^2}\right)^2}\,\cos (k\pi)
  =0\; ,
\end{equation}
where the $\pm$ sign corresponds to $\bar\omega_2\gtrless 0$.
The parameters $\bar\nu$ and $\bar B$ determine the
solution of this implicit equation for the variables $\bar q$ and $|\bar\omega
|$.
The solution gives these specific values of the viscosity
for which either positive or negative purely imaginary growth rates exist
(see Fig.\ \ref{fig:1}). For a supercritical induction of $\bar B=1.05$
and $\bar q=1$ there exist a positive purely imaginary growth rate for {\it
all}
viscosities ($k=0$). Above a critical viscosity, $\bar\nu_c =0.453$, a negative
purely imaginary growth rate solves (\ref{eq:49}) as well. This critical
viscosity also
gives the upper bound for the solution with $k=1$. The value of
$\bar\nu_c =\bar\nu_c (\bar q, \bar B)$ increases with increasing induction
at constant $\bar q$ and decreases with increasing wave vectors at constant
$\bar B$.
The critical viscosity is naturally zero at the onset of the instability,
$\bar\nu_c (\bar q =1, \bar B =1) =0$.

\begin{figure}[htbp]
  \begin{center}
    \includegraphics[scale=0.45]{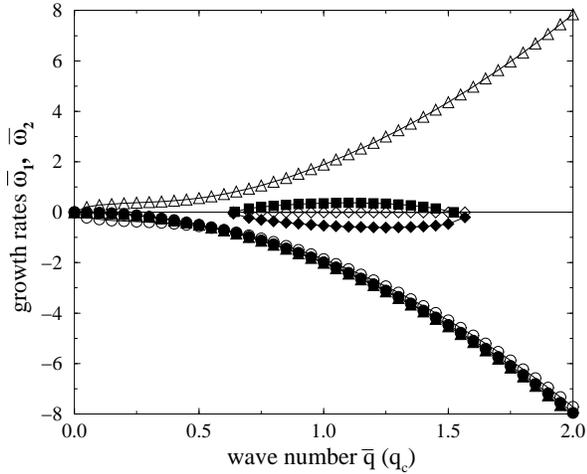}
    \caption{Growth rate $\bar\omega =\bar\omega_1 +i\,\bar\omega_2$ as
    function of the wave number
    $\bar q$ for $\bar\nu=0.037$ and $\bar B=1.05$. The empty symbols
    ($\scriptstyle{\protect\triangle}$, $\diamond$, $\circ$) show $\Re
    (\bar\omega )=\bar\omega_1$
    and the filled symbols ($\blacktriangle$,
    $\scriptstyle{\protect\blacklozenge}$,
    $\bullet$, $\scriptstyle{\protect\blacksquare}$)
    display $\Im (\bar\omega )=\bar\omega_2$. A positive purely imaginary
    growth rate ($\scriptstyle{\protect\blacksquare}$) exists only in the
    vicinity of $\bar q=1$.}
    \label{fig:2}
  \end{center}
\end{figure}

\begin{figure}[htbp]
  \begin{center}
    \includegraphics[scale=0.45]{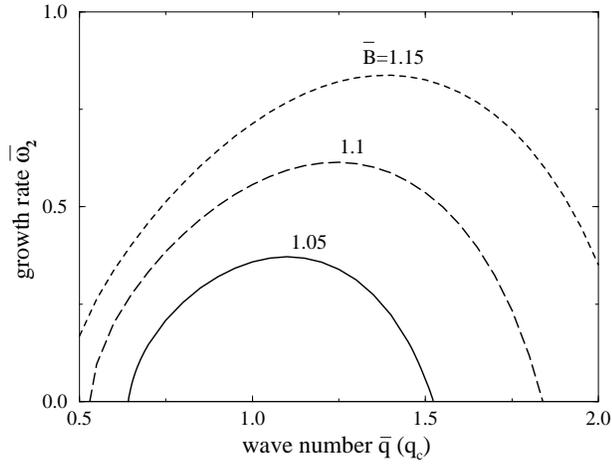}
    \caption{Positive purely imaginary growth rates as a function of the wave
    number $\bar q$ and the
    strength of the induction $\bar B$ for $\bar\nu=0.037$, typically for
    magnetic fluids in experiments
    \protect\cite{mahrdiss,mahr98}. The maximum is given by $\bar\omega_{2,m}$
    and $\bar q_m$ which both increase monotonously with $\bar B$.}
    \label{fig:3}
  \end{center}
\end{figure}

Whereas Fig.\ \ref{fig:1} shows a situation where certain types of solutions of
the dispersion relation exist, in Fig.\ \ref{fig:2} the complete solution of
Eqs.\
(\ref{eq:46}, \ref{eq:47}) for $\bar B=1.05$ and
$\bar \nu=0.037$ is plotted ($k=0$). Around the critical wave number $\bar
q_c=1$ a
range of wave numbers exist with positive purely imaginary growth rates (filled
squares),
i.e., there is a band of unstable wave vectors. All other growth rates have
negative
imaginary parts (filled circles, filled diamonds, and filled triangles).
Therefore the unstable wave
vectors are related to the positive purely imaginary growth rates only.
Focusing on
this solution, the imaginary part $\bar \omega_2$ is shown in the vicinity of
$\bar q_c$
for various strengths of the induction and $\bar \nu=0.037$
($\nu\simeq 6.4\cdot 10^{-6}\,{\rm m}^2{\rm s}^{-1}$) in Fig.\ \ref{fig:3}. The
chosen value of the
viscosity characterizes typical magnetic fluids in experiments
\cite{mahrdiss,mahr98}.
All three curves have a maximum in the growth rate $\bar\omega_m =i\,\bar
\omega_{2,m}$
at $\bar q_m$. One notes that $\bar \omega_{2,m}$ as well as $\bar q_m$ are
monotonously increasing functions of the strength of the supercritical
induction
at constant viscosity.

\begin{figure}[htbp]
  \begin{center}
    \includegraphics[scale=0.45]{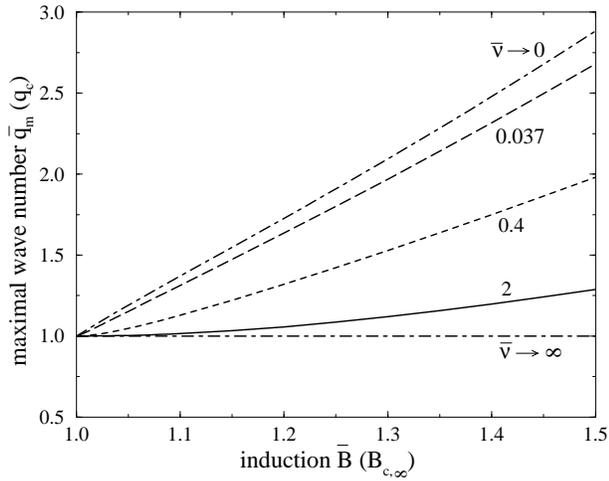}
    \caption{Maximal wave number $\bar q_m$ as function of the supercritical
    induction $\bar B$ for different viscosities. $\bar q_m$ is a monotonously
    increasing function of $\bar B$ with the exception $\bar q_m=1$
    \protect\cite{salin93} in the case of infinitely large
    viscosities (lower dot-dashed line). In the limit of an inviscid fluid
    (upper dot-dashed line) the dependence of $\bar q_m$ on $\bar B$ is
    given by $\bar q_m = (1/3)\left( 2\bar B^2+\sqrt{4\bar B^4-3}\;\right)$
    \protect\cite{abou97}.}
    \label{fig:4}
  \end{center}
\end{figure}

In order to study the resulting behaviour of
$\bar\delta\bar q =\bar q\sqrt{2\bar\nu /|\bar\omega |}$
at $\bar q_m$, the details of the dependence of $\bar \omega_{2,m}$ and $\bar
q_m$ on
the field {\it and} the viscosity need to be known. The wave number with
the maximal growth rate is defined by $\partial \bar\omega_2/\partial \bar q =
\partial |\bar\omega |/\partial \bar q =0$. Since $|\bar\omega |$ is given
implicitly
by $f_\pm (\bar q , |\bar\omega |;{\bar\nu ,\bar B})=0$, the maximal growth
rate
results from
\begin{eqnarray}
  \nonumber
  g(\bar q , |\bar\omega |;{\bar\nu ,\bar B}) &:=&{\partial f_+\over \partial
  \bar q} =0\\
  \nonumber
    &=& 256\bar\nu^5\left[ 1-\cos^2(k\pi)\right] \bar q^8
    +96\bar\nu^3\bar q^7 +\left\{ 9\bar\nu +128\left[ -\bar\nu^3\bar B^2+
    \left( 4+3 \cos^2(k\pi)\right)\bar\nu^4|\bar\omega |\right]\right\}\bar
    q^6\\
  \nonumber
    &&+\,8\left( -3\bar\nu\bar B^2 +4\bar\nu^3 +18\bar\nu^2 |\bar\omega
    |\right)\bar q^5
    +\left\{ 16\left[ 20 -9\cos^2(k\pi)\right] \bar\nu^3 |\bar\omega |^2
    +6\bar\nu
    +16\bar\nu\bar B^4 +9|\bar\omega | -192\bar\nu^2|\bar\omega |\bar
    B^2\right\}\bar q^4\\
  \nonumber
    &&+\,8\left( 6\bar\nu^2|\bar\omega | +6\bar\nu |\bar\omega |^2
    -3|\bar\omega |\bar B^2
    -\bar\nu \bar B^2\right)\bar q^3 +\left( -64\bar\nu |\bar\omega |^2\bar B^2
    +64\bar\nu^2|\bar\omega |^3 +6|\bar\omega | +16|\bar\omega |\bar B^4
    +\bar\nu\right)\bar q^2\\
  \label{eq:50}
    &&+8\left( 2\bar\nu |\bar\omega |^2 -|\bar\omega |\bar B^2\right)\bar q +
    |\bar\omega |\; .
\end{eqnarray}
The cross section of the solutions of Eqs.\ (\ref{eq:49}, \ref{eq:50}) gives
$|\bar\omega_m|$
and $\bar q_m$, which is shown for three different viscosities in Fig.\
\ref{fig:4}.
Besides a viscosity of real magnetic fluids, two large viscosities ($\bar\nu
=0.4$, $2$) were
chosen to represent the regime where the behaviour of the fluid is dominated by
the
viscosity.
For all three viscosities the wave number $\bar q_m$ is {\it not constant}, i.e., for finite
viscosities $\bar q_m$ depends on the external control parameter $\bar B$.
With increasing viscosity $\bar q_m$ varies
less with increasing induction, e.g. $\Delta\bar q_m = 1.68$ for $\bar\nu
=0.037$
reduces to $\Delta\bar q_m = 0.29$ for $\bar\nu =2$ at an induction  difference
of
$\Delta\bar B=0.5$. For small viscosities $\bar q_m$ depends linearly on $\bar
B$ if
$\bar B$ is not too large. This linear dependence shifts towards higher values
of $\bar B$
with increasing viscosity (compare $\bar\nu =0.037$ and $0.4$). At the largest
viscosity,
$\bar\nu =2$, no linear behaviour can be observed for $1 <\bar B <1.5$.

\begin{figure}[htbp]
  \begin{center}
    \includegraphics[scale=0.45]{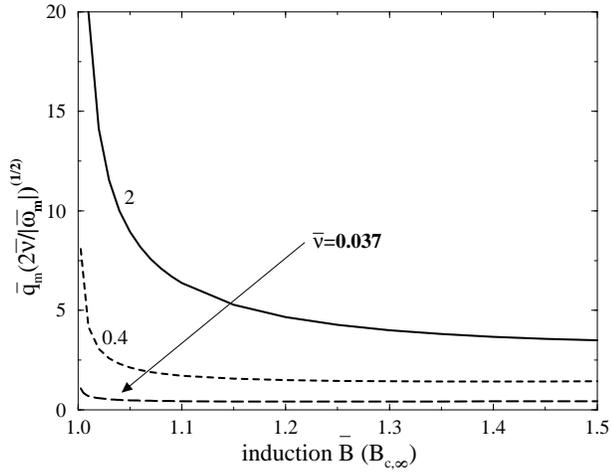}
    \caption{Behaviour of $\bar q\bar\delta$ at $\bar q_m$ as a function of the
    induction
    for the viscosities $\bar\nu=0.037$, $0.4$,
    and $2$. The condition $\bar q_m\bar\delta\gg 1$ holds only in the close
    vicinity of the critical induction, $\bar B \approx 1$, for large
    viscosities
    ($\bar\nu =0.4$, $2$) and in the limit $\bar B =1$ for realistic
    viscosities
    ($\bar\nu =0.037$).}
    \label{fig:5}
  \end{center}
\end{figure}

The analysis reveals that only in the case of infinitely large viscosities
(with respect
to the viscosity of real magnetic fluids) a constant wave vector of maximal
growth
$\bar q_m=1$ can be expected.
Taking into account viscous effects does not necessarily lead to a constant
$\bar q_m$.
For a better comparison with \cite{salin93} the value of $\bar q\bar\delta$ at
$\bar q_m$
is calculated and is plotted for the three viscosities chosen in Fig.\
\ref{fig:5}.
The graphs show clearly that $\bar q_m\bar\delta\gg 1$ holds only in the close
vicinity of the critical induction for large viscosities
($\bar\nu =0.4$, $2$) and in the limit $\bar B= 1$
($|\bar\omega_m |= 0$, $\bar\delta =\infty$) for realistic viscosities
($\bar\nu =0.037$).
Because $0.43<\bar q_m\bar\delta <1.08$ for $\bar\nu =0.037$, realistic fluid
properties
are neither covered by the other asymptotic regime $\bar q_m\bar\delta\ll 1$
analysed
in \cite{salin93}. Therefore the experimental observation in
\cite{cowley,bacri84} can
not be explained by the result of an asymptotic analysis which does not meet
the features
of the experimental fluids. By plotting the know analytical result in the
inviscid regime, $\bar q_m = (1/3)\left( 2\bar B^2+\sqrt{4\bar B^4-3}\right)$
\cite{abou97}, realistic magnetic fluids tend rather to the limit
$\bar\nu\rightarrow 0$ than to the limit $\bar\nu\rightarrow\infty$
(see Fig.\ \ref{fig:4}) as exploited in earlier studies \cite{cowley}.
But for {\it quantitative} comparisons in typical experimental setups, asymptotic
analyses \cite{salin93,abou97} are insufficient.

\section{Finite Layer of Magnetic Fluid}

\begin{figure}[htbp]
  \begin{center}
    \includegraphics[scale=0.45]{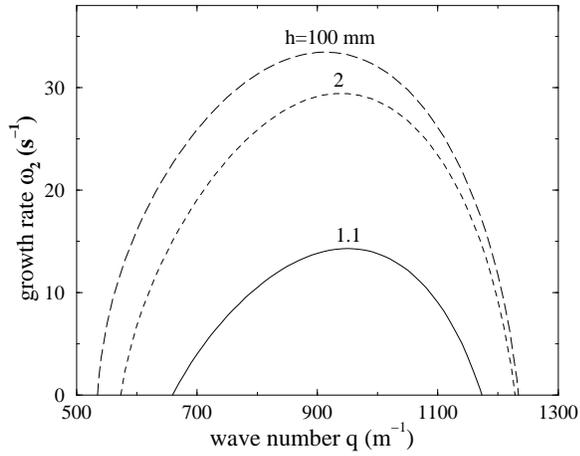}
    \caption{Positive purely imaginary growth rates as function of the wave
    number $q$ and the thickness
    of the layer at a constant induction of $B=106\cdot 10^{-4}$ T. Remarkably
    the graph for a layer of $2$ mm is
    already close to the limit of an infinite thick layer illustrated by
    $h=100$ mm. Whereas the wave number
    of maximal growth shows only a small variation, the maximal growth rate
    itself displays more distinct
    changes. Material parameters of the fluid EMG 901 are listed in Table
    \ref{table1}.
    }
    \label{fig:6}
  \end{center}
\end{figure}

Since the experiments are performed with a vessel of finite depth, the method
presented
in the previous section has to be applied to magnetic fluids of finite
thickness.
With the polar representation of $\omega$, the real and imaginary
part
of the dispersion relation (\ref{eq:12}) read
\begin{eqnarray}
  \label{eq:60}
  &&\nu^2\left[ {R_1\over N_1}\left( {R_2\over N_2} - R_3 - {R_4\over
  N_4}\right)
  +{I_1\over N_1}\left( {I_2\over N_2} - I_3 -{I_4\over N_4}\right) \right]
  + \tanh (qh) \left[ gq +{\sigma\over \rho}q^3-{(\mu_r -1)^2 B^2\over
  \mu_0\,\mu_r\,\rho}
   \Lambda (qh)\,q^2\right]  =  0\\
  \label{eq:61}
  &&\nu^2\left[ {R_1\over N_1}\left( {I_2\over N_2} - I_3 - {I_4\over
  N_4}\right)
  +{I_1\over N_1}\left( -{R_2\over N_2} + R_3 +{R_4\over N_4}\right) \right]  =
  0\, .
\end{eqnarray}
The explicit form of the abbreviations $R_i$, $I_i$, $N_1$, $N_2$, and $N_3$
($i=1, \ldots ,4$) is
deferred to Appendix A. For purely imaginary growth rates, $\omega =i\omega_2$,
Eq.~(\ref{eq:61}) is fulfilled without any restrictions for $\omega_2 >0$ as
well as for
$\omega_2 <0$ in contrast to the case of an infinitely thick layer. Only
positive purely
imaginary growth rates are of interest for comparison with the experiment.
Therefore
the function $f$ is now of the form
\begin{eqnarray}
  \nonumber
  &&f_+(q,\omega ;\nu, B, h):={\nu^2\left[\cosh (2\tilde q_1 h) -1\right]\over
    \tilde q_1\sinh (2\tilde q_1 h)-q\coth (qh)\left[\cosh (2\tilde q_1 h)
    -1\right] }
  \left\{ {\tilde q_1\sinh (2\tilde q_1 h)(5q^4 +2q^2 \tilde q_1^2 +\tilde
  q_1^4)\over
  \cosh (2\tilde q_1 h) -1}\right.\\
  \label{eq:62}
  &&\left. -q \tanh (qh)\left( 5q^2\tilde q_1^2 +q^4 +\tilde q_1^4\right)
  -{4q^2 \tilde q_1 (q^2+\tilde q_1^2)\over \cosh (qh) \sinh (\tilde q_1
  h)}\right\}
  +\tanh (qh) \left[ gq +{\sigma\over \rho}q^3-{(\mu_r -1)^2 B^2\over
  \mu_0\,\mu_r\,\rho}
   \Lambda (qh)\,q^2\right]=0\; ,
\end{eqnarray}
where this implicit equation for $q$ and $\omega$ contains the additional
parameter $h$.
Fig.\ \ref{fig:6} shows the solution for three different depths of the layer at
a supercritical
induction of $B=106\cdot 10^{-4}$ T. The used material parameters of the
commercially available
magnetic fluid EMG 901 (Ferrofluidics Corporation) are listed in Table
\ref{table1}.
The graphs show that the wave number of maximal growth $q_m$ clearly
varies less with $h$ than the maximal growth rate $\omega_{2,m}$ itself.
Notably,
the solution for a layer of $2$ mm thickness is already near the infinite case
illustrated by $h=100$ mm. Therefore a filling with $h \geq 1$ cm  of magnetic
fluid can be considered as an infinite thick layer. To make such an estimate
is an asset in the use of the complete equations. Because they cover the entire
range of thickness, $0\leq h \leq \infty$, in extension to the asymptotic
analysis $h\simeq 0$ and $h\simeq\infty$ in \cite{abou97}.

\begin{figure}[htbp]
  \begin{center}
    \includegraphics[scale=0.45]{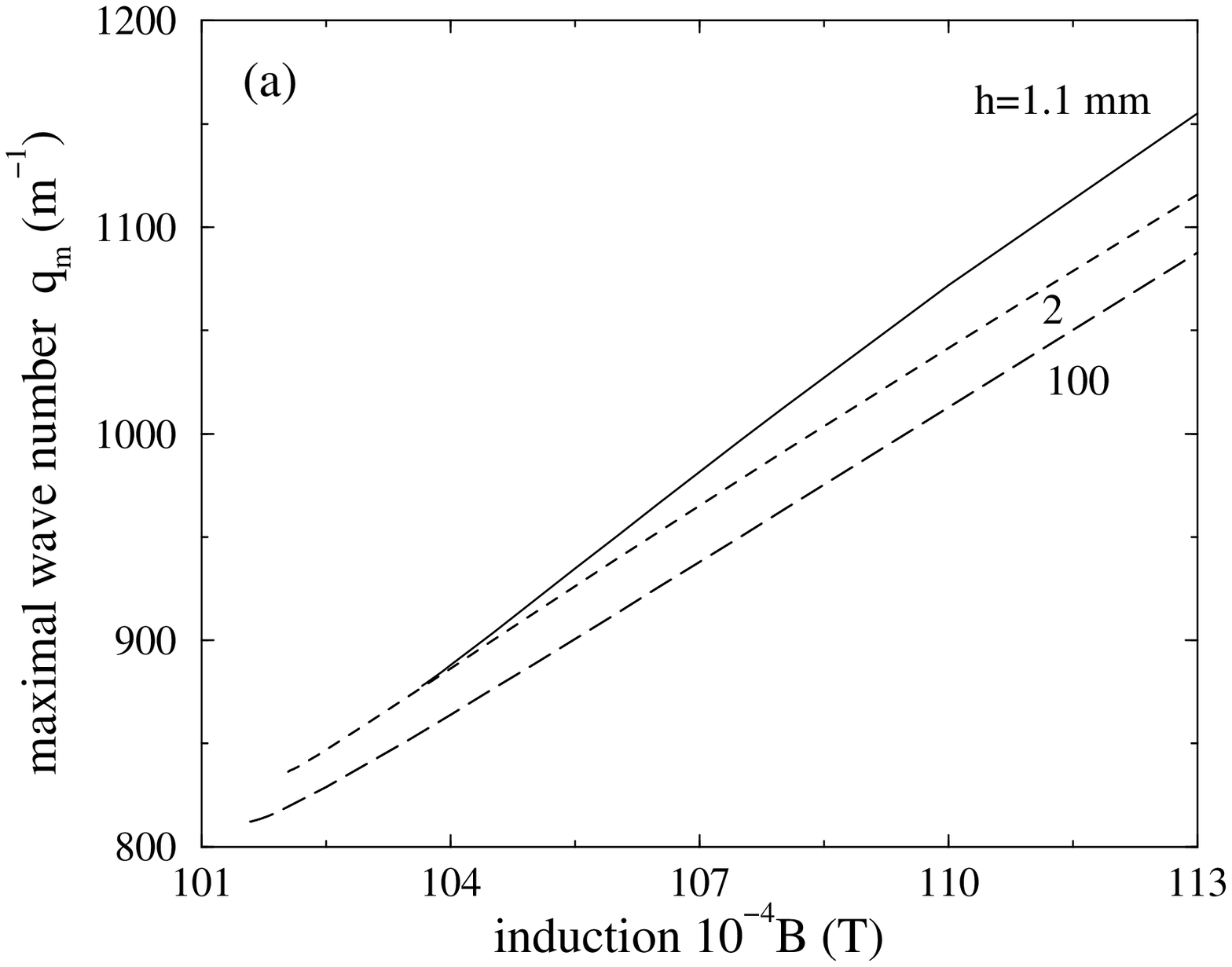}
    \includegraphics[scale=0.45]{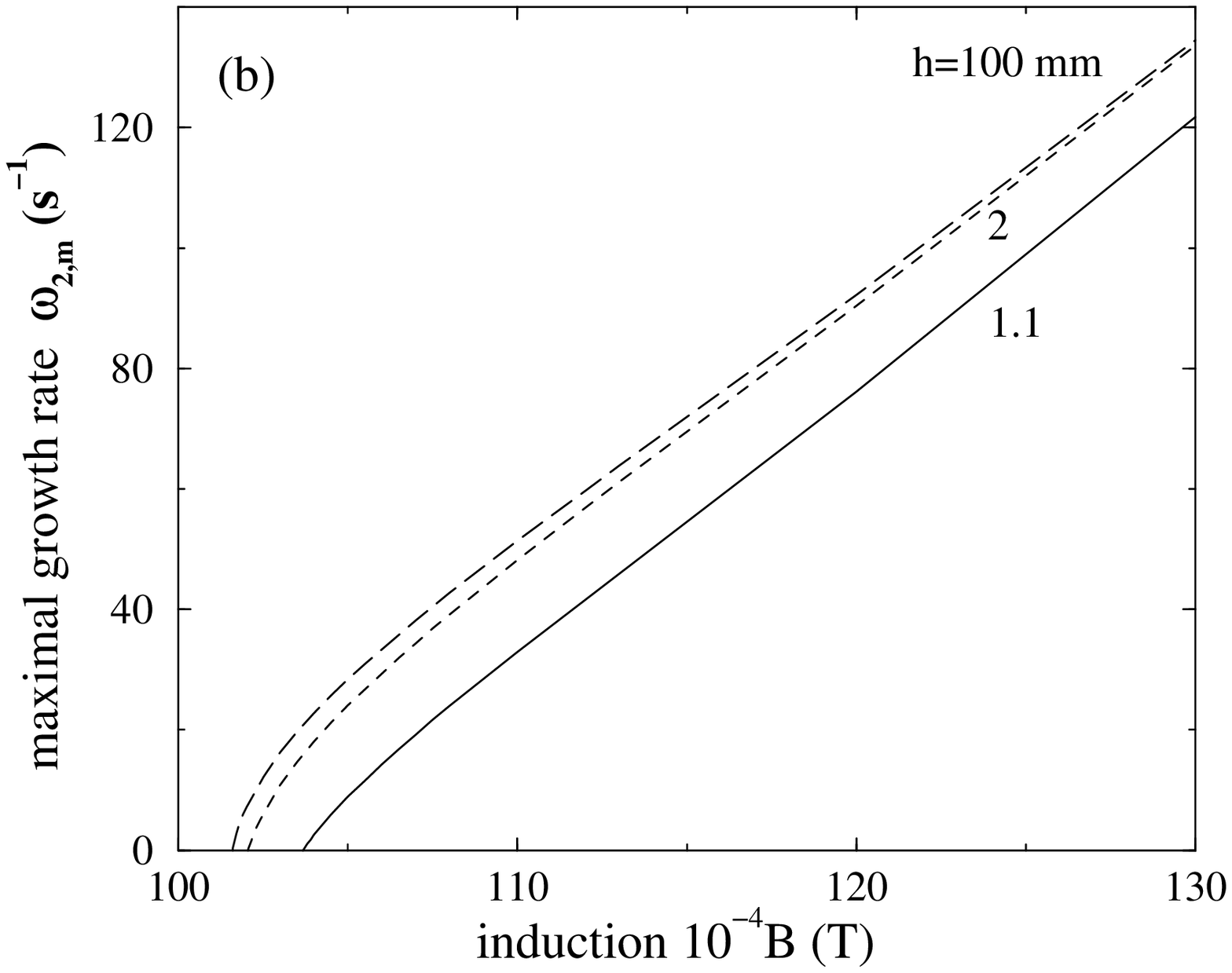}
    \caption{Maximal wave number $q_m$ (a) and maximal growth rate
    $\omega_{2,m}$ (b) as function
     of the supercritical induction $B>B_{c,h}$ for three different
     thicknesses.
     (a) $q_m$ increases
     linearly with $B$ except for $B$ near the critical value $B_{c,h}$. The
     area of nonlinear behaviour
     shrinks with the shrinking thickness of the layer. (b) $\omega_{2,m}$
     starts to grow like a square
     root above the onset of the instability. This square root behaviour
     becomes less pronounced with
     thinner layers. Material parameters of the fluid EMG 901 are listed in
     Table \ref{table1}.
     }
    \label{fig:7ab}
  \end{center}
\end{figure}

To analyse the behaviour of $\omega_{2,m}$ and $q_m$ on $B$ and $h$, the
maximal growth
rate, given by $\partial f_+/\partial q=0$, has to be determined. As the
resulting implicit
function is quite lengthy thus we do not give the explicit form here.
The cross section of the solutions of Eq.~(\ref{eq:62}) and $\partial
f_+/\partial q=0$
leads to $\omega_{2,m}$ and $q_m$. Their dependence on the supercritical
induction and
the thickness of the layer is shown in Fig.\ \ref{fig:7ab}. The wave number of
maximal
growth increases linearly with $B$ with the exception of $B$ near the
height dependent critical value $B_{c,h}$.
The linear behaviour is independent of the thickness of the layer and
holds up to $30\%$ above $B_{c,h}$. The maximal growth rate
$\omega_{2,m}$ starts to grow like a square root
above the onset of the instability. This square root behaviour
becomes less pronounced with thinner layers.

\begin{figure}[htbp]
  \begin{center}
    \includegraphics[scale=0.45]{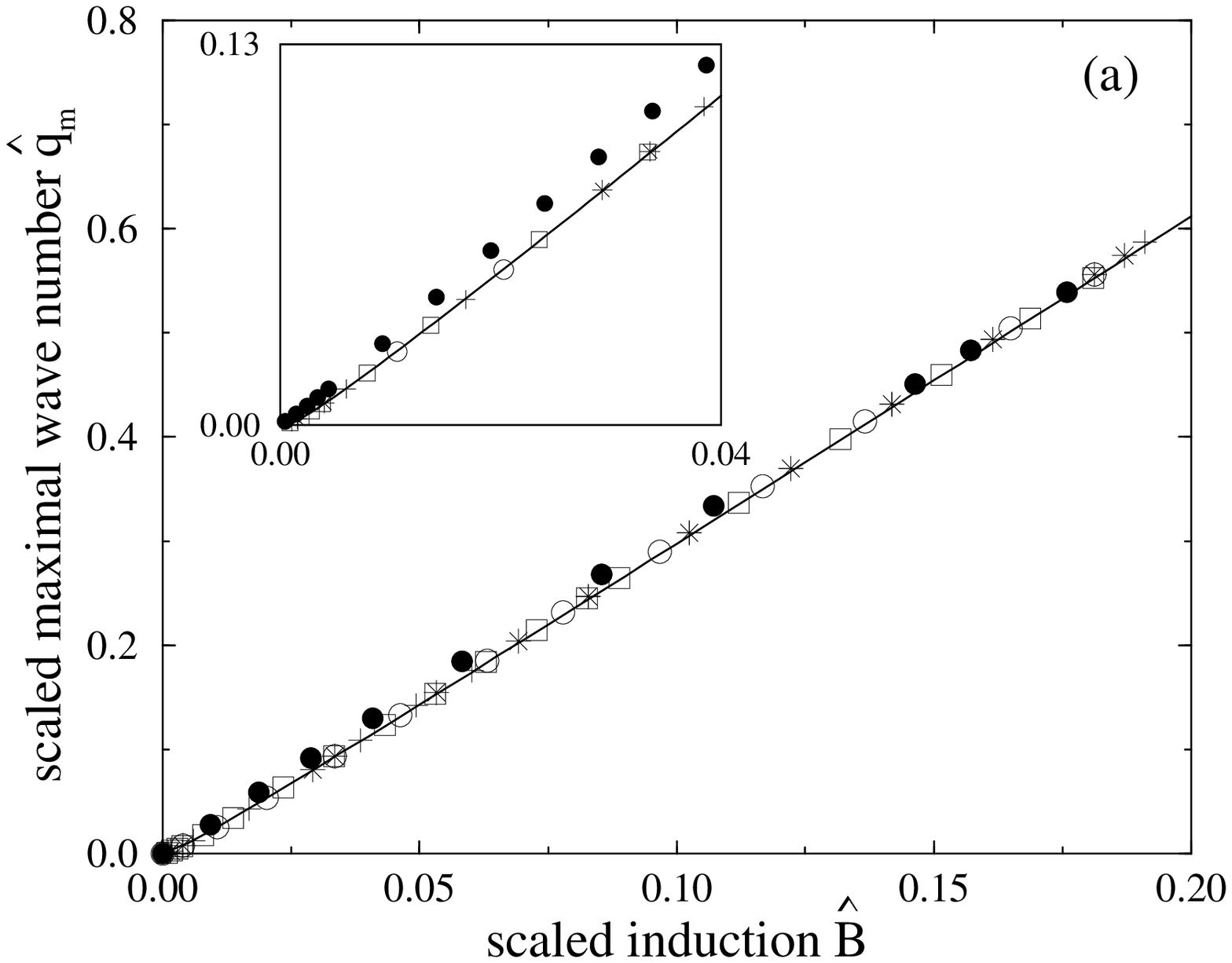}
    \includegraphics[scale=0.45]{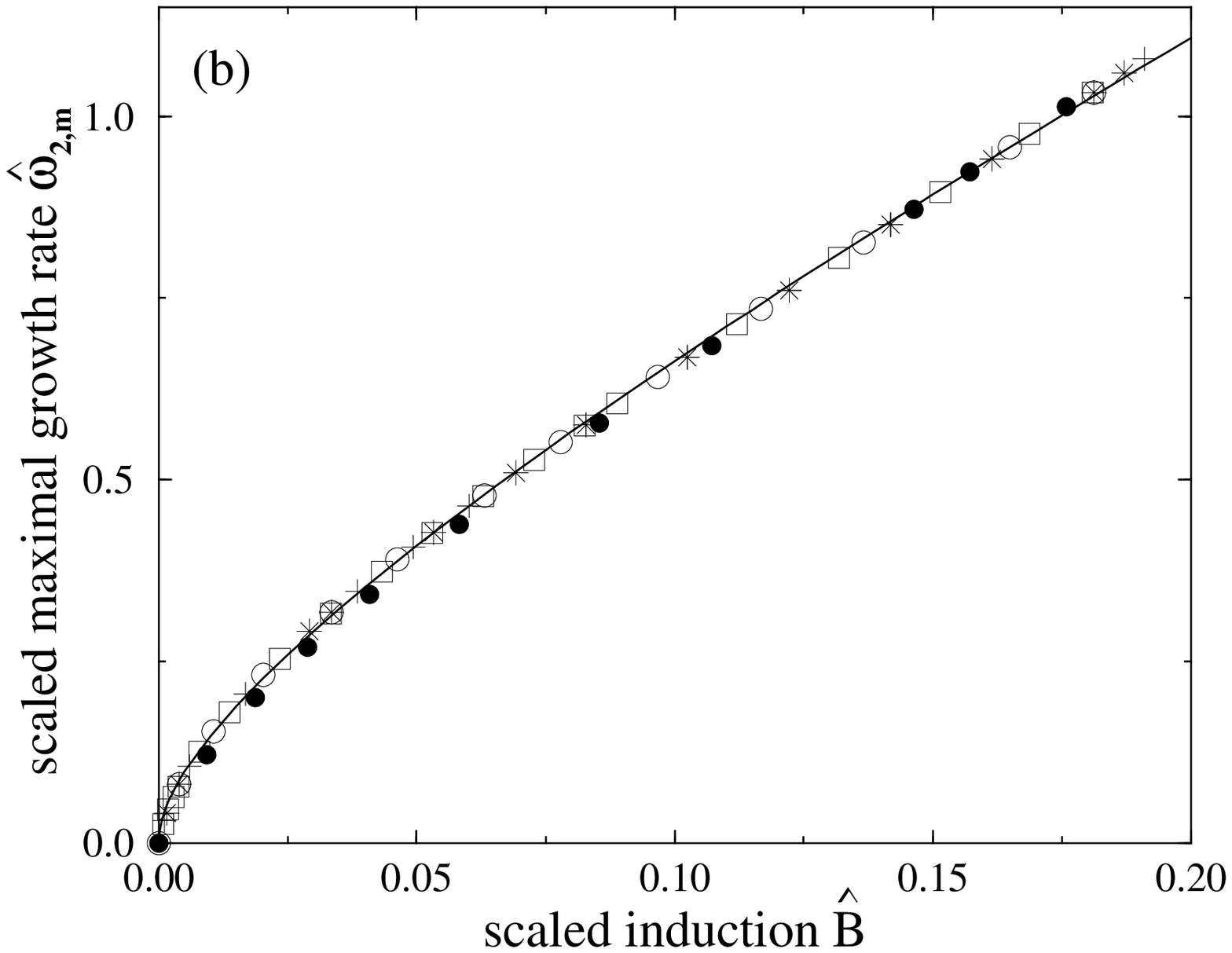}
    \caption{Scaled maximal wave number $\hat q_m$ (a) and scaled maximal
    growth rate
     $\hat \omega_{2,m}$ (b) as function of the scaled supercritical
     induction $\hat B$. The data are calculated for $h=100$ mm ($\circ$), $50$
     mm ($\ast$),
     $10$ mm ($+$), $4$ mm ($\scriptstyle{\square}$), $2$ mm ($\bullet$).
     For $h\geq 4$ mm the data are fitted by $\hat q_m =3.26 \hat B
     -0.09\sqrt{\hat B}$ for
     $\hat q_m$ (solid line (a)) and by $\hat \omega_{2,m}  = 1.18\sqrt{\hat B}
     + 2.9\hat B$
     for $\hat \omega_{2,m}$ (solid line (b)). Small deviations from the
     generic
     behaviour can be seen for $h=2$ mm (insert). Material parameters of the
     fluid EMG 901
     are listed in Table \ref{table1}.}
    \label{fig:7cd}
  \end{center}
\end{figure}

Through the implicit character of the functions an analytical expression cannot
be given
for the dependence of $q_m$ and $\omega_{2,m}$ on $B$ and $h$. Alternatively, a
two parameter
fit is tested, which describes the generic behaviour over a wide range of layer
thicknesses.
An excellent agreement is achieved for $h\geq 4\,{\rm mm}$ by
\begin{eqnarray}
  \label{eq:63}
  \hat q_m &=&3.26 \hat B -0.09\sqrt{\hat B}\quad{\rm for~}0.001\leq \hat B
  \leq 0.2\\
  \label{eq:64}
  \hat \omega_{2,m} & =& 1.18\sqrt{\hat B} + 2.9\hat B \;\;\quad{\rm
  for~}0.001\leq \hat B \leq 0.2\; ,
\end{eqnarray}
(see Fig.\ \ref{fig:7cd}) where $\hat B= (B-B_{c,h})/B_{c,h}$,
$\hat q_m =(q_m-q_{c,h})/q_{c,h}$, and $\hat \omega_{2,m}=\omega_{2,m} t_c$
denote the scaled distances from the critical values.
For {\it small} $\hat B$ the behaviour of $\hat q_m$ is only weakly nonlinear
whereas the behaviour of $\hat \omega_{2,m}$ is determined by the square root
term.
A careful inspection of the data reveals that for $h=2$ mm (filled circles)
small deviations from the proposed fits appear: $\hat q_m$ grows linearly
over the entire $\hat B$ region (see insert in Fig.\ \ref{fig:7cd} (a)). Thus
$h=2$ mm
indicates the lower limit of the validity of (\ref{eq:63}, \ref{eq:64}).

\begin{figure}[htbp]
  \begin{center}
    \includegraphics[scale=0.45]{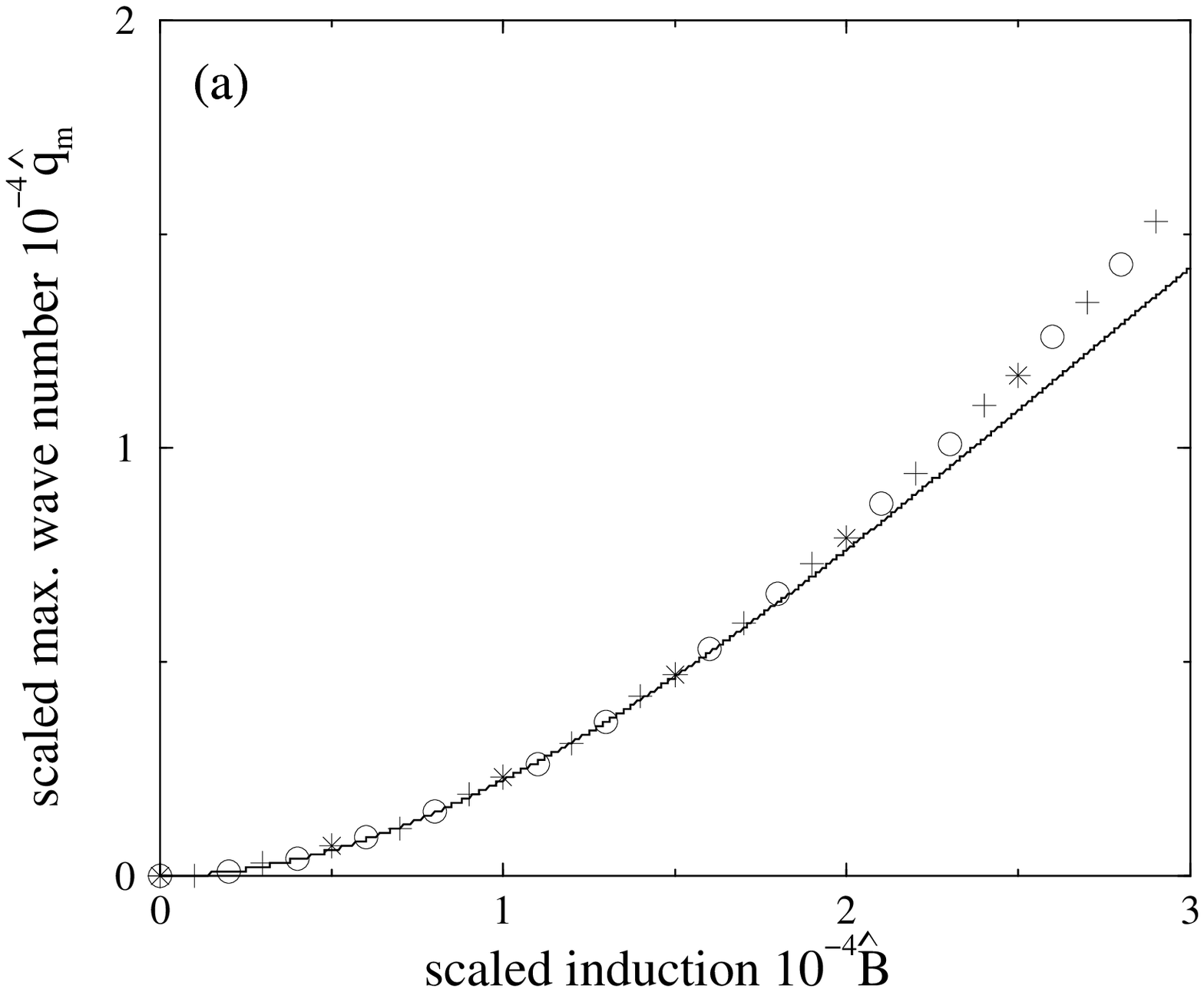}
    \includegraphics[scale=0.45]{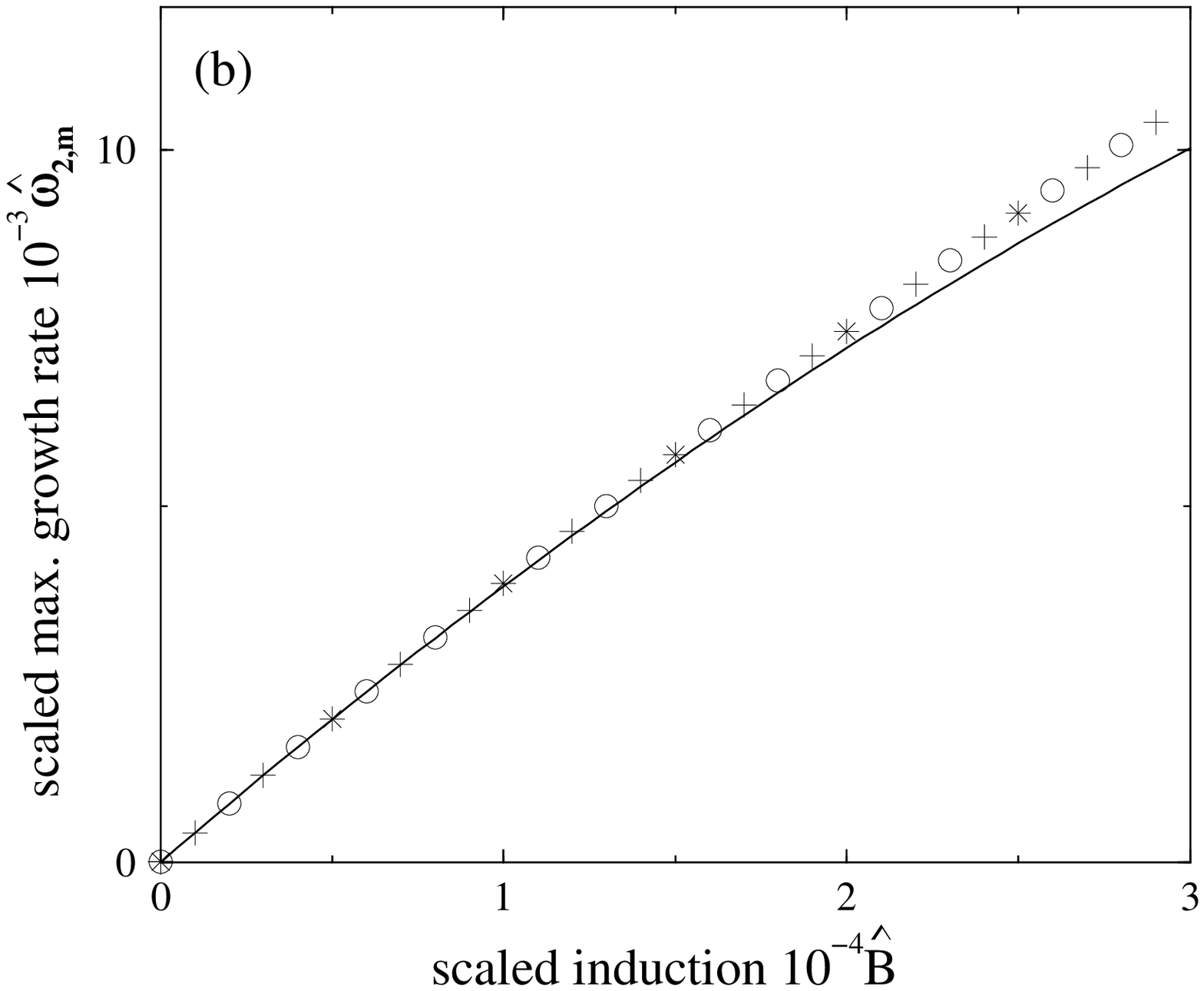}
    \caption{Scaled maximal wave number $\hat q_m$ (a) and scaled maximal
    growth rate
     $\hat \omega_{2,m}$ (b) as function of the scaled supercritical
     induction $\hat B$. The data are calculated for $h=100$ mm ($\circ$), $50$
     mm ($\ast$), and
     $10$ mm ($+$). For $\hat B< \bar\nu^2/6$, i.e. $\hat B<4\cdot 10^{-4}$ for
     EMG 901, the agreement
     with the analytical
     results $\hat q_m = (6/\bar\nu^2)\hat B^2 - (18/ \bar\nu^4)\hat B^3$
     (solid line (a)) and
     $\hat \omega_{2,m}  =(2/\bar\nu)\hat B - (3/ \bar\nu^3 )\hat B^2$ (solid
     line (b)) is very good.
     Material parameters of the fluid EMG 901 are listed in Table
     \ref{table1}.}
    \label{fig:7ef}
  \end{center}
\end{figure}

Since the fit covers the region of infinite thick layers one can expand
(\ref{eq:49}, \ref{eq:50}) for small $\hat B$, $\hat q_m$, and $\hat
\omega_{2,m}$.
Taking into account that the dimensionless viscosity is
also small for real magnetic fluids ($\bar\nu\simeq 0.0483$ for EMG 901)
the expansion leads to
\begin{eqnarray}
  \label{eq:65}
  \hat q_m &=& {6\over \bar\nu^2}\hat B^2 - {18\over \bar\nu^4 }\hat B^3
  \quad{\rm for~}0\leq \hat B \ll {\bar\nu^2\over 6}\\
  \label{eq:66}
  \hat \omega_{2,m} & =&{2\over \bar\nu}\hat B - {3\over \bar\nu^3 }\hat B^2
  ~~~\quad{\rm for~}0\leq \hat B \ll {\bar\nu^2\over 6}\; ,
\end{eqnarray}
where the coefficients depend on the viscosity. Fig.\ \ref{fig:7ef} shows the
very good
agreement between the numerical solution and the expansion for $\hat B \ll
\bar\nu^2/6$.
The region where the expansion holds extends with the square of the viscosity.
The expansion
(\ref{eq:65}, \ref{eq:66}) and the scaling (\ref{eq:63}, \ref{eq:64}) show that
the
behavior of $\hat q_m$ and $\hat \omega_{2,m}$ is entirely governed by the two
parameters
viscosity and induction for not too thin layers. The third parameter $h$ has
only little
effect in this regime.

\begin{figure}[htbp]
  \begin{center}
    \includegraphics[scale=0.45]{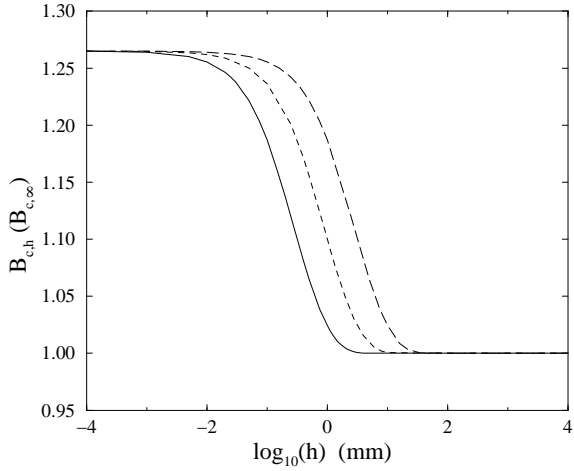}
    \caption{The critical induction $B_{c,h}$ versus the thickness of the layer
    $h$ for three
    different surface tensions: $\sigma=2.275\cdot 10^{-2}\,{\rm kg}\,{\rm
    s}^{-2}$ (solid line),
    $\sigma=2.275\cdot 10^{-1}\,{\rm kg}\,{\rm s}^{-2}$ (dashed line), and
    $\sigma=2.275\,{\rm kg}\,{\rm s}^{-2}$ (long-dashed line).
    By increasing the surface tension by a factor of $10$ ($100$), $B_{c,0}$
    can be measured for
    layers nearly 1 (2) orders of magnitude thicker than for a system with the
    original surface
    tension. The remaining material parameters of the fluid EMG 901 are
    listed in Table \ref{table1}.
    }
  \label{fig:10}
 \end{center}
\end{figure}

From the results shown in Fig.\ \ref{fig:7ab} one notes the height dependence
of
$B_c$ and $q_c$ at the onset of the instability (see also Fig.\ 4, 5
in\cite{weilepp}).
This dependence for $B_c$ can be exploited to measure the permeability of the
magnetic fluid `just in time' for the experiment. Since the quotient of the
two limits (\ref{eq:15}, \ref{eq:16}) depends on $\mu_r$ only
\begin{equation}
  \label{eq:67}
  {B_{c,0}\over B_{c,\infty}} = {2\mu_r\over \mu_r +1}\; ,
\end{equation}
the determination of the two limits of $B_c$ offers a feasible access to
$\mu_r$ of the
magnetic fluid. From Fig.\ \ref{fig:10} it can be seen that $B_c$ increases
monotonously
from $B_{c,\infty}$ towards $B_{c,0}$ with decreasing layer thickness.
Since the preparation of very thin layers is laborious and delicate it would be
desirable to shift the thin layer limit towards thicker films. This can be
achieved by an
increase of the surface tension. A modified surface tension is
accompanied by changes in the density and permeability of the fluid. But these
changes
are of a much smaller scale than those of the surface tension. The modified
viscosity does
not affect the determination of $B_{c,h}$. Therefore the surface tension is
changed
whereas all other quantities remain constant. By increasing the surface tension
by a
factor of $10$ ($100$), $B_{c,0}$ may be measured for layers nearly 1 (2)
orders of magnitude
thicker than for a system with the original surface tension (see Fig.\
\ref{fig:10}).

\begin{figure}
   \includegraphics[scale=0.45]{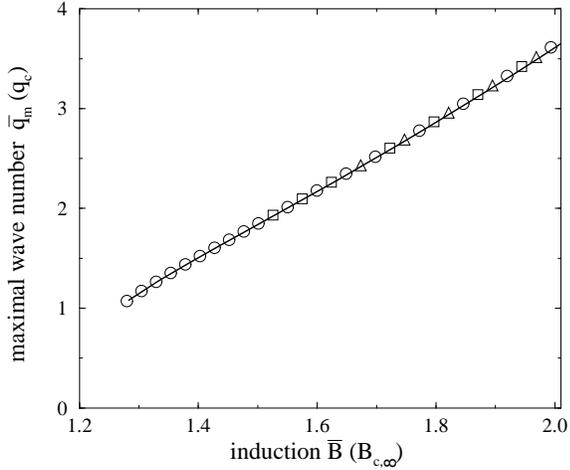}
   \caption{Maximal wave number $\bar q_m$ as function of the supercritical
   induction $\bar B$ of a thin film, $h=1\mu{\rm m}$, for
   three different viscosities: $\nu = 6.5\cdot 10^{-8}\,{\rm m}^2\,{\rm s}^{-1}$
   ($\circ$), $\nu = 6.5\cdot 10^{-6}\,{\rm m}^2\,{\rm s}^{-1}$
   ($\square$), and $\nu = 2.0\cdot 10^{-5}\,{\rm m}^2\,{\rm s}^{-1}$
   ($\triangle$). The numerical data show that the behaviour of $\bar q_m$
   on $\bar B$ is independent of $\nu$ and it is given by
   $\bar q_m = (1/4)\left( c + \sqrt{c^2-8}\,\right)$
   with $c=3\bar B^2 (\mu_r +1)/(2\mu_r)$ \protect\cite{abou97} (solid line).
   The remaining material parameters of the fluid EMG 901 are
   listed in Table \ref{table1}.}
   \label{thinfilm}
\end{figure}

Abou et al. analysed the limit of thin films of magnetic fluid for vanishing
and infinitely large viscosities. In both cases
the analytical result \cite{abou97}
\begin{equation}
\bar q_m = {1\over 4}\left( 3\bar B^2{(\mu_r + 1)\over 2\mu_r}+
\sqrt{9\bar B^4 {(\mu_r + 1)^2\over 4\mu_r^2}-8}\;\right)
\end{equation}
is the same. Therefore, one can assume that the dependence of $\bar q_m$
on $\bar B$ is not influenced by the viscosity in the thin film limit. Since
the present method allows to calculate $\bar q_m$ for any
combination of parameters, we are able to accomplish a test of this
assumption. As Fig. \ref{thinfilm} shows, the behaviour of $\bar q_m$
on $\bar B$ is indeed independent of $\nu$ (The tested viscosities
cover a range from $6.5\cdot 10^{-8}\,{\rm m}^2\,{\rm s}^{-1}$ to
$2\cdot 10^{-5}\,{\rm m}^2\,{\rm s}^{-1}$.). The variation of $\bar q_m$
on the applied magnetic induction in thin films was measured in earlier
experiments where the magnetic fluid either was prepared at the bottom of a
quartz chamber \cite{skjeltorp} or was laid on top of a denser fluid
\cite{bacri88,petit92}. In all three experiments the spacing between
the {\em final} arrangement of peaks was measured in dependence of the
applied field and a nonlinear behaviour was found.

\section{Measurement, Results, and Comparison with Theory}

In this section we report on  experimental results of the dependence of
the maximal wave number on the supercritical magnetic induction.
First we present the experimental setup, next we give a characteristic example
of the pattern evolution. We continue with a description of the techniques
applied to extract the wave number of the patterns. Finally the experimental
results are compared with the theoretical results of the previous section,
particularly the predicted growth of the maximal wave number.

\begin{figure}
   \epsfxsize=8cm\epsfbox{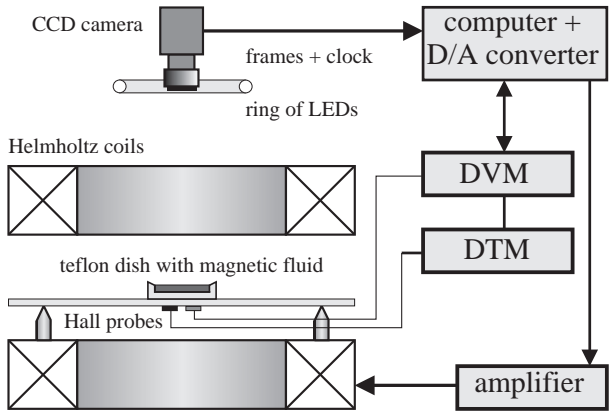}\medskip
   \caption{
   Scheme of the experimental setup
   }
   \label{setup}
\end{figure}

Our experimental setup is shown in Fig.~\ref{setup}. A cylindrical Teflon$^{\rm
\circledR}$ vessel with a diameter of $\rm d=12~cm$ and a depth of 3~mm is
completely filled with magnetic fluid and situated in the center of a pair of
Helmholtz coils. The experiments were performed with EMG 909. The fluid is
illuminated by 90 red LEDs mounted on a ring of 30~cm diameter placed in a
distance of 105~cm above the surface. A CCD-camera is positioned at the center
of the ring. By this construction a flat fluid surface reflects no light into
the camera lens, however, an inclined surface of proper angle will reflect
light into the camera~\cite{bechhoefer1995}. The CCD-camera is connected via a
framegrabber to a Pentium 90 MHz PC and serves additionally as a fundamental
clock for timing the experiment. In the theoretical analysis the supercritical
magnetic field is assumed to be instantly present, thus in the experiment the
magnetic field has to be increased jump-like from a subcritical value $B_0$ to
the desired value $B$. For all measurements $B_0$ was fixed to
$133\cdot10^{-4}\,\rm T$. The jump-like increase of the field is initiated by
the computer. Its D/A converter is connected via an amplifier (fug Elektronik
GmbH) to the Helmholtz coils (Oswald Magnetfeldtechnik). The magnetic system
can not follow the control signal instantly, its relaxation time $\tau_B$ to a
jump--like increase of the control signal depends on the jump hight $\Delta B
=B-B_0$. For a maximal jump hight of $\Delta B=70\cdot10^{-4}\,{\rm T}$ the
relaxation time mounts up to $\tau_B=80\,\rm ms$. The other characteristic time
scales of the system are the capillary time scale, $t_c\simeq 13\,\rm ms$, and
the viscous time scale, $t_\nu =1/(q_c^2\, \nu) \simeq 450\,\rm ms$.

For the empty Helmholtz--coils the spatial homogeneity of the magnetic field is
better than $\pm 1 \%$. This grade is valid within a cylinder of 10~cm in
diameter and 14~cm in hight oriented symmetrically around the center of the
coils. Two Hall-probes are positioned immediately under the Teflon$^{\rm
\circledR}$ dish. A Siemens Hall-probe (KSY 13) serves to measure the magnetic
field during its jump--like increase, and is connected to the digital voltmeter
(Prema 6001). For measuring a constant magnetic field and for calibration
purpose we use a commercial Hall-probe (Group3-LPT-231) connected to the
digital teslameter (DTM 141). Both devices are controlled via IEEE-bus by the
computer.

\begin{figure}
   \epsfxsize=16cm\epsfbox{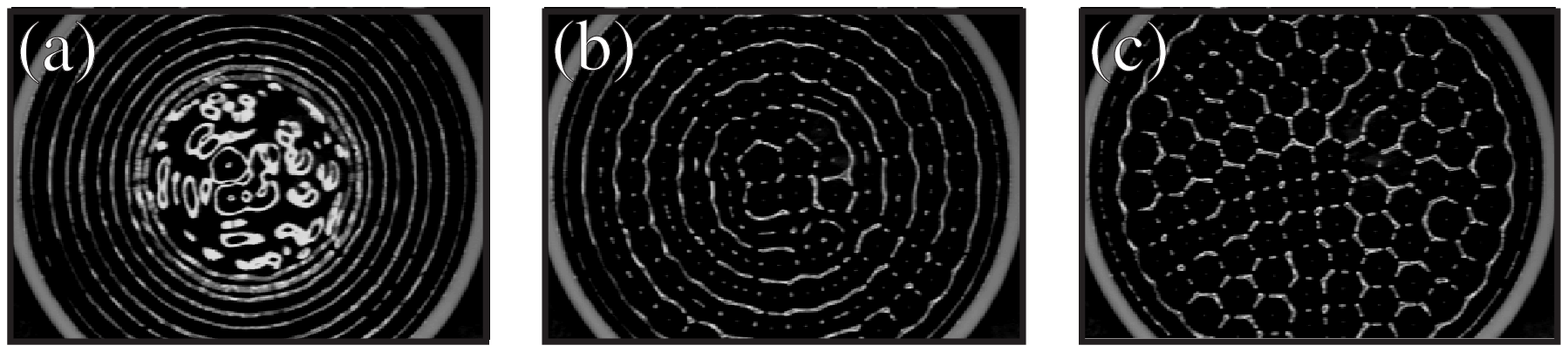}\medskip
   \caption{
   Series of snap-shots of the principal pattern evolution of the
   magnetic fluid for a jump from $B_0<B_c$ to $B>B_c$, illuminated from above by
   a ring of LEDs. The pictures are taken $\Delta t=180$ ms (a), $280$
   ms (b), $560$ ms (c) after the start of the increase of the magnetic field.
   }
   \label{evolution}
\end{figure}

Next we give a characteristic example for the evolution of the surface pattern
during a jump--like increase of the magnetic field. Fig.~\ref{evolution}(a)
shows circular surface deformations taken $\Delta t=180\,\rm ms$ after the
start of the experiment. This surface deformations are first created at the
edge of the dish, because of the discontinuity of the magnetic induction
induced by the finite size of the container. The circular deformation is fixed
in space, and its amplitude grows during the jump--like increase of the
magnetic field. With increasing time more circular deformations evolve,
approaching the center of the dish (see Fig.~\ref{evolution}(a)). Onto this
pattern, Rosensweig peaks emerge out of the crests of the circular surface
deformation, as can be seen in Fig.~\ref{evolution}(b). After this transient
concentric arrangement a hexagonal pattern of Rosensweig peaks evolves (see
Fig.~\ref{evolution}(c)).

The theoretical results stem from a linear theory which can determine correctly
the critical values of the pattern selected by the instability only at the
threshold. Above the threshold, a band of wave numbers will become unstable,
where the mode with the largest growth rate is the most unstable linear mode of
the flat interface. Due to nonlinear effects the final stable pattern does not
generally correspond to the most unstable linear mode, as shown here and in
other experiments \cite{godreche}. Therefore it has to be stressed that for the
comparison with the linear theory not the stable hexagonal pattern, but the
most early stage of the pattern, namely the transient circular deformations are
most appropriate.

\begin{figure}
   \epsfxsize=8cm\epsfbox{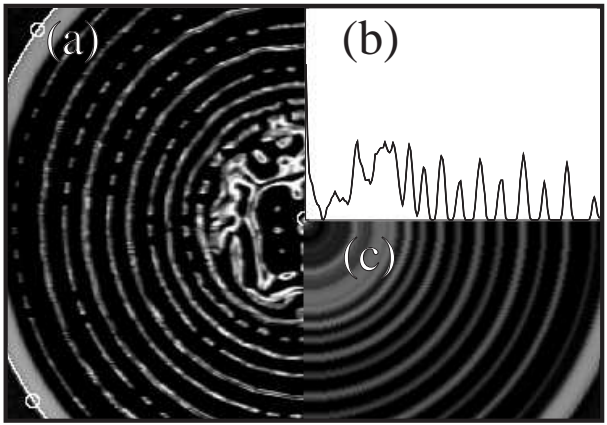}\bigskip
   \caption{
   Three steps of the picture processing to extract the wave number: (a)
   Reflections of circular surface deformations. The white line in the left
   upper and
   lower corner marks the calculated edge of the dish, the circles on the edge
   serve to calculate the center of the dish. (b) Radial grey level
   distribution of (a). (c) Two dimensional representation of (b).
   }
   \label{pic_proc}
\end{figure}

The wave number of the circular deformations is extracted from the pictures in
the following way. First the algorithm scans the diagonals of the picture for
the local maximum of the grey levels. Starting at the corners of the pictures,
it detects points which are situated at the edge of the Teflon$^{\rm
\circledR}$ dish. Two of the edge points are marked by white circles in the
left part of Fig.~\ref{pic_proc}(a). From the full set of four points the
algorithm calculates the center of the dish denoted by the half circle at the
right part of Fig.~\ref{pic_proc}(a). In order to control the precision of the
algorithm a white circle with the proper radius of the dish is constructed
around the detected center. In the next step we calculate the radial
distribution of the grey values for all pixels within the circumference of the
dish, as shown in Fig.~\ref{pic_proc}(b). For comparison
Fig.~\ref{pic_proc}(c) gives an artificial two dimensional representation of
the grey value distribution.

In all of the three pictures of Fig.~\ref{pic_proc} one can easily discriminate
three zones. In the innermost zone only small surface undulations exist, which
give rise to the unstructured part of the grey value distribution in
Fig.~\ref{pic_proc}(b,c). For larger radii one finds the area of circular
surface deformations which  generates the bi--periodic peak pattern in the
distribution. The large peaks are correlated with the deformation troughs and
the small peaks with the deformation crests. Finally the outermost zone
includes the edge of the dish together with the first, edge induced,
deformation crest. For estimation of the wave number we discard the inner- and
outermost zones. The top part of the peaks in the remaining zone is fitted by a
polynomial of second order. The average distance of their maxima gives the half
of the desired wavelength. For the rather small number of deformations
available, the above presented method of wave number calculation turned out to
be more stable and to give more precise results than the competing method of
two dimensional Fourier transformation. Together with the picture the
momentarily magnetic induction has been recorded during the jump--like
increase. This allows an exact relation of the extracted wave number to the
instantly prevailing magnetic induction.

\begin{figure}
   \epsfxsize=8cm\epsfbox{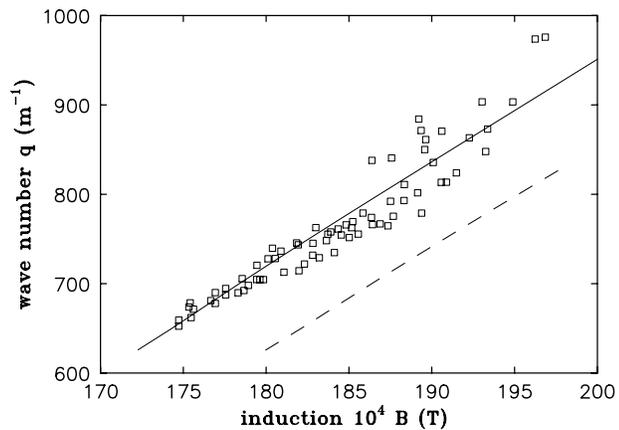}\medskip
   \caption{
   Plot of the wave number $q$ versus the magnetic induction $B$. The
   open squares give the experimental values, the dashed line displays the
   theoretical results for the material parameters of EMG 909 listed in
   Table \ref{table1}. Using $\mu_r$ as a fit--parameter gives the solid line.
   }
   \label{ex_vs_th}
\end{figure}

Let us now focus on the experimental results displayed in Fig.~\ref{ex_vs_th},
where the wave number $q$ is plotted versus the magnetic induction $B$. Each
open square denotes the wave number extracted from a picture taken during a
jump--like increase of the magnetic field to $B>B_c$. The estimated maximal
errors for $q$ of $\pm 4.2\%$ and for $B$ of $\pm 0.9\%$ are not plotted for
the purpose of clarity. The dashed line displays the theoretical results for
the listed material parameters of the magnetic fluid EMG~909. Using $\mu_r$ as
a fit--parameter gives the solid line with $\mu_r\simeq 1.86$. The fitted value
for  $\mu_r$ differs by $3.3\%$ from the value given by Ferrofluidics, a
deviation which is well within the tolerance of production specified by
Ferrofluidics. Obviously there is a rather good agreement between the
experimental results and the theoretical graph for $\mu_r\simeq 1.86$ marked by
the solid line. The {\em linear} increase in the appearing wave number, both in
experiment and in theory, is our main outcome.

Comparing the two theoretical curves in Fig.~\ref{ex_vs_th}, an increase of
$\mu_r$ results in a decrease of the critical magnetic induction, whereas the
critical wave number remains constant, as can bee seen from Eqs.~(\ref{eq:15},
\ref{eq:16}, \ref{eq:17}). According to Eq.~(\ref{eq:17}) a constant critical
wave number implies that the density and the surface tension are constant.
Therefore we refrain to consider them as additional fit parameters.
As can be seen from Fig.~\ref{fig:4} changes in
the viscosity by an order of magnitude are necessary to cause a relevant
influence on the behaviour of wave number of maximal growth on the induction.
Therefore changes in the viscosity due to small thermal fluctuations in the
experiment can be neglected.

We find a linear wave number dependence of the circular surface deformations.
This pattern is a more simple realization of the normal field instability than
the familiar hexagonal pattern of Rosensweig cusps. The latter one is obtained
by a symmetrical superposition of {\em three} patterns of parallel stripes with
the wave vectors separated by 120 degree \cite{cross1993}. Obviously the
circular surface deformations can be regarded as a stripe pattern favoured by
the symmetry of the dish. As a consequence they appear first, before nonlinear
interactions select in a later stage the hexagonal pattern. This situation is
well known from Rayleigh-Benard convection in cylindrical containers, where,
due to side-wall induced convection, concentric target patterns appear instead
of hexagonal structures. Our observations agree in part with recent findings by
Browaeys et al.~\cite{browaeys}. They detected circular surface deformations
for a constant, subcritical magnetic field of $0.79 H_c$. In contrast to their
experiment, we do not perform a periodic modulation, but a jump--like increase
of the magnetic induction. Thus we have no interference with additional waves
propagating onto the circular deformations. Therefore a measurement of the wave
number, as described above, could be realized.

The circular surface deformations have to be distinguished from circular,
meniscus induced surface waves emitted from the edge of lateral cell
walls~\cite{douady}. Here, the circular deformations are induced by the
discontinuity of the magnetic induction at the edge of the container. The
formation of a meniscus is eluded by a brimful filling of the dish and by the
design of the vessel which has a slope with respect to the horizontal of
$15^\circ$, the contact angle between the magnetic fluid and Teflon$^{\rm
\circledR}$.

Finite size effects due to the finite size of the vessel are rather small in
the experiment. Applying the arguments of Edwards and
Fauve \cite{edwards94}, the width of the band of unstable wave numbers, $\Delta
q\simeq 5.1$ cm$^{-1}$ for $B=180\cdot10^{-4}\,{\rm T}$ and $h=3$ mm, is much
larger than $\pi/d\simeq 0.3$ cm$^{-1}$, the wave number separation between the
quantized modes of the vessel. Thus the influence of the vessel size can be
neglected and the developing pattern is insensitive to the vessel size.

For the experiments we have chosen a magnetic fluid with
a rather low value of magnetic permeability $\mu_r =1.86$,
in order to keep hysteresis effects small. Indeed, with
our resolution a hysteresis can not be detected.
Hysteresis strength proves to increase monotonically with
the permeability of the magnetic fluid \cite{boudouvis}.
Thus, the influence of higher permeability on the wave number of maximal
growth remains to be investigated experimentally in the context of a
more complex situation of a transition with large hysteresis.

\section{Summary}
By means of the polar representation of the complex frequency $\omega$
the dispersion relation for surface waves on viscous magnetic fluids is split
into a real and an imaginary part. The parameters are determined for which pure
imaginary solutions $\omega =i\omega_2$ and $\omega_2 >0$ for both parts exist.
For these parameters the originally horizontal surface is unstable, because
initially small undulations of the surface, proportional to $\exp (-i\omega
t)$, grow exponentially. The imaginary part of the dispersion relation is
fulfilled mostly automatically. From the real part, the wave number with
maximal growth rate $q_m$ and the maximal growth rate $\omega_{2,m}$ itself can
be easily determined. It can be done for any combination of material parameters
and for any thickness of the layer. This is the strength of the presented
analytical method which covers the entire parameter space between the previously
studied asymptotic cases \cite{salin93,abou97}. It therefore allows to study
the transition from one limit to the other. Such a transition is exemplarily
illustrated for an infinitely thick layer with viscosities varying between zero and
infinity. 

For magnetic fluids of infinite depth it is shown that earlier qualitative
observations of constant wave numbers above the critical magnetic field
\cite{cowley,bacri84} cannot be explained by the result of an asymptotic
analysis \cite{salin93}. The analysis in \cite{salin93} does not cover the
features of the experimental fluids. In order to apply a theory, where the
field is instantly present, a jump--like increase of the field in the
experiments is essential. Therefore the results for a continuously increased
field \cite{cowley,bacri84} are inappropriate for a comparison with such a
theory. Furthermore, we have demonstrated that the transient pattern is the
most suitable one to be compared to the linear theory. Taking all these into
account, we are able to observe a linear increase of the wave number
of maximal growth with increasing
magnetic induction. This linear increase is quantitatively confirmed by the
linear theory.

An increasing wave number with increasing field was also observed in the
corresponding electrical setup where a liquid metal is subjected to a normal
electric field \cite{desurgy}, but this result is based only on a qualitative
observation. The authors emphasize as well the importance of a fast build up of
the field.

It is very attractive to test in further experiments whether the predicted
generic behaviour of the maximal growth rate can be confirmed in the weakly
nonlinear regime. As an expected outcome, $\omega_{m,exp}$ should start to grow
like a square root with increasing supercritical induction. Furthermore, it
remains to be seen whether the linear increase of the wave number of the
linearly most unstable pattern lasts in the final hexagonal pattern. A
confirmation would mean that also the wave number of the final pattern varies
if the induction is {\em jump-like} increased. We point out, that for a {\em
continuous} increase of the induction the behaviour of the wave number for both
the transient and the final pattern remains to be elucidated.

\section*{Acknowledgment}

The authors profit from stimulating discussions with Johannes Berg, Andreas
Engel,
Ren\'e Friedrichs, Hans Walter M\"uller, and Ingo Rehberg. This work was
supported
by the Deutsche Forschungsgemeinschaft under Grant EN 278/2.

\section{Appendix}

The abbreviations in Eqs.\  (\ref{eq:60},\ref{eq:61}) read explicitly
\begin{eqnarray}
  \label{eq:80}
  R_1 &=& [\cosh (2\tilde q_1 h) -\cos (2\tilde q_2 h)] \left\{ \tilde q_1\sinh
  (2\tilde q_1 h)
          +\tilde q_2 \sin (2\tilde q_2 h) -q\coth (qh)\left[ \cosh (2\tilde
          q_1 h) -\cos (2\tilde q_2 h)
          \right]\right\}\, ,\\
  \label{eq:81}
  I_1 &=& [\cosh (2\tilde q_1 h) -\cos (2\tilde q_2 h)] \left[ \tilde q_2\sinh
  (2\tilde q_1 h)
          -\tilde q_1\sin (2\tilde q_2 h)\right] \, ,\\
  \label{eq:82}
  N_1 &=& \left\{ \tilde q_1\sinh (2\tilde q_1 h) +\tilde q_2 \sin (2\tilde q_2
  h) -q\coth (qh)
          \left[ \cosh (2\tilde q_1 h) -\cos (2\tilde q_2 h) \right]\right\}^2
          +\left\{ \tilde q_2\sinh (2\tilde q_1 h) -\tilde q_1\sin (2\tilde q_2
          h)\right\}^2 \, ,\\
  \nonumber
  R_2 &=& \left[ 5q^2+2q^2(\tilde q_1^2 -\tilde q_2^2) +\tilde q_1^4 -6 \tilde
  q_1^2\tilde q_2^2
          +\tilde q_2^4 \right] \left[ \tilde q_1\sinh (2\tilde q_1 h)  +
          \tilde q_2 \sin (2\tilde q_2 h) \right]\\
  \label{eq:83}
      & & -\left[ 4 q^2 \tilde q_1\tilde q_2 +4 \tilde q_1^3
          \tilde q_2 -4\tilde q_1\tilde q_2^3\right] \left[ \tilde q_2\sinh
          (2\tilde q_1 h)  -
          \tilde q_1 \sin (2\tilde q_2 h) \right]\, ,\\
  \nonumber
  I_2 &=& \left[ 4 q^2 \tilde q_1\tilde q_2 +4 \tilde q_1^3 \tilde q_2 -4\tilde
  q_1\tilde q_2^3\right]
          \left[ \tilde q_1\sinh (2\tilde q_1 h)  + \tilde q_2 \sin (2\tilde
          q_2 h) \right]\\
  \label{eq:84}
      & & +\left[ 5q^2+2q^2(\tilde q_1^2 -\tilde q_2^2) +\tilde q_1^4 -6 \tilde
      q_1^2\tilde q_2^2
          +\tilde q_2^4 \right] \left[ \tilde q_2\sinh (2\tilde q_1 h)  -
           \tilde q_1 \sin (2\tilde q_2 h) \right]\, ,\\
  \label{eq:85}
  N_2 &=& \cosh (2\tilde q_1 h) -\cos (2\tilde q_2 h)\, ,\\
  \label{eq:86}
  R_3 &=& q\tanh (qh)\left[ 6q^2 (\tilde q_1^2 -\tilde q_2^2) +q^4 +\tilde
  q_1^4
          -6\tilde q_1^2\tilde q_2^2 + \tilde q_2^4\right]\, , \\
  \label{eq:87}
  I_3 &=& q\tanh (qh)\left[ 12q^2 \tilde q_1\tilde q_2 +4\tilde q_1^3\tilde q_2
          - 4\tilde q_1\tilde q_2^3\right]\, ,\\
  \label{eq:88}
  R_4 &=& 4q^2 \sinh (\tilde q_1 h)\cos (\tilde q_2 h)\left[ \tilde q_1 (q^2
  +\tilde q_1^2
          -\tilde q_2^2) -2\tilde q_1\tilde q_2^2\right] +4q^2\cosh (\tilde q_1
          h)\sin (\tilde q_2 h)
          \left[ \tilde q_2 (q^2 +\tilde q_1^2 -\tilde q_2^2) +2\tilde
          q_1^2\tilde q_2\right]\, ,\\
  \label{eq:89}
  I_4 &=& 4q^2 \sinh (\tilde q_1 h)\cos (\tilde q_2 h)\left[ \tilde q_2 (q^2
  +\tilde q_1^2
          -\tilde q_2^2) +2\tilde q_1^2\tilde q_2\right]- 4q^2\cosh (\tilde q_1
          h)\sin (\tilde q_2 h)
          \left[ \tilde q_1 (q^2 +\tilde q_1^2 -\tilde q_2^2) -2\tilde
          q_1\tilde q_2^2\right]\, ,\\
  \label{eq:90}
  N_4 &=& \cosh (qh) \left[ \cosh^2 (\tilde q_1 h)-\cos ^2(\tilde q_2
  h)\right].
\end{eqnarray}
In (\ref{eq:80}-\ref{eq:90}) the shorthands
\begin{eqnarray}
  \label{eq:91}
  \tilde q_1 &=& \sqrt[\scriptstyle{4}]{\left(q^2+{|\omega |\sin\varphi_0 \over
  \nu}
                 \right)^2 +{|\omega |^2 \cos^2 \varphi_0\over \nu^2}}\;
                 \cos\left({\psi +2 k\pi\over 2}\right)\, ,\\
  \label{eq:92}
  \tilde q_2 &=& \sqrt[\scriptstyle{4}]{\left(q^2+{|\omega |\sin\varphi_0 \over
  \nu}
                 \right)^2 +{|\omega |^2 \cos^2 \varphi_0\over \nu^2}}\;
                 \sin\left({\psi +2 k\pi\over 2}\right)
\end{eqnarray}
were used where
\begin{equation}
   \label{eq:93}
   \psi = {\rm arctan}\,{-|\omega |\cos\varphi_0\over q^2\nu + |\omega |\sin
   \varphi_0}
          + {\rm const}\, .
\end{equation}

\narrowtext

\eject

\widetext

\narrowtext

\widetext
\begin{table}
\caption{Material parameters of EMG 901 and EMG 909.}
\label{table1}
\begin{tabular}{ldcdc}
\noalign{\smallskip}
                             & EMG 901 & Source & EMG 909 & Source  \\
\noalign{\smallskip}\hline\noalign{\smallskip}
$\mu_r$                      &  4.0                  & Ferrofluidics
& 1.8                 & Ferrofluidics\\
$\rho$ [kg$\cdot$m$^{-3}$]   &  1.53$\cdot 10^{3}$   & \cite{rothert} &
1.02$\cdot 10^{3}$  & Ferrofluidics\\
$\nu$ [m$^2\cdot$s$^{-1}$]   &  6.54$\cdot 10^{-6}$  & Ferrofluidics
& 5.88$\cdot 10^{-6}$ & Ferrofluidics\\
$\sigma$ [kg$\cdot$s$^{-2}$] &  2.27$\cdot 10^{-2}$  & \cite{rothert}   &
2.65$\cdot 10^{-2}$ & \cite{mahrdiss}\\
\noalign{\smallskip}
\end{tabular}
\end{table}


\begin{references}
\bibitem{comment1}
Since $\omega$ is generally a complex number, 
the absolute value of $\omega$ should appear in the definition of the viscous depth,
$\delta =\sqrt{2\nu /|\omega (\nu, q)|}$.
\end{references}
\end{document}